\documentclass{article}
\usepackage[T1]{fontenc} 
\usepackage[utf8]{inputenc} 
\usepackage{ismir,amsmath,cite,url}
\usepackage{graphicx}
\usepackage{color}
\usepackage{booktabs}
\usepackage{multirow}
\usepackage{lineno}
\usepackage{bm}
\usepackage{algpseudocode}
\usepackage{algorithm}   
\usepackage[bookmarks=false]{hyperref}

\usepackage[dvipsnames]{xcolor}
\newcommand\myshade{70}
\colorlet{mywholecolor}{MidnightBlue}
\hypersetup{
  linkcolor  = mywholecolor!\myshade!black,
  citecolor  = mywholecolor!\myshade!black,
  urlcolor   = mywholecolor!\myshade!black,
  colorlinks = true,
}

\newcommand\dotsubsec[1]{\vspace{0.8mm}\noindent\textbf{#1. }}

\newcommand{\fref}[1]{Figure~\ref{#1}}
\newcommand{\tref}[1]{Table~\ref{#1}}
\newcommand{\sref}[1]{Section~\ref{#1}}

\def\vx{{\bm{x}}}

\def\vz{{\bm{z}}}

\title{ST-ITO: Controlling audio effects for \\ style transfer with inference-time optimization}

\multauthor
{Christian J. Steinmetz \hspace{0.5cm} Shubhr Singh \hspace{0.5cm} Marco Comunità \hspace{0.5cm} Ilias Ibnyahya} { \bfseries{Shanxin Yuan \hspace{0.5cm} Emmanouil Benetos \hspace{0.5cm} Joshua D. Reiss} \\
Centre for Digital Music, Queen Mary University of London, UK \\
\vspace{-0.5cm}
}

\def\authorname{C. J. Steinmetz, S. Singh, M. Comunità, I. Ibnyhahya, S. Yuan, E. Benetos, and J. D. Reiss}

\begin{document}

\maketitle
\begin{abstract}
Audio production style transfer is the task of processing an input to impart stylistic elements from a reference recording. Existing approaches often train a neural network to estimate control parameters for a set of audio effects. However, these approaches are limited in that they can only control a fixed set of effects, where the effects must be differentiable or otherwise employ specialized training techniques. In this work, we introduce \mbox{\textbf{ST-ITO}}, Style Transfer with Inference-Time Optimization, an approach that instead searches the parameter space of an audio effect chain at inference. This method enables control of arbitrary audio effect chains, including unseen and non-differentiable effects. Our approach employs a learned metric of audio production style, which we train through a simple and scalable self-supervised pretraining strategy, along with a gradient-free optimizer. Due to the limited existing evaluation methods for audio production style transfer, we introduce a multi-part benchmark to evaluate audio production style metrics and style transfer systems. This evaluation demonstrates that our audio representation better captures attributes related to audio production and enables expressive style transfer via control of arbitrary audio effects.

\end{abstract}
\section{Introduction}\label{sec:intro}


Audio effects are signal processing devices used to transform or manipulate audio signals, such as adding reverberation, adjusting frequency balance with equalization, or adding edge with distortion. 
They play a central role in audio production, providing audio engineers with the ability to realize both practical and creative tasks with applications in music, film, broadcast, and video games~\cite{wilmering2020history}.
Traditionally, operating these effects requires a significant amount of expertise, as audio engineers must combine a technical understanding with their artistic goals. 
As a result, the process of creating a high-quality audio production remains challenging, requiring a time-consuming process for professionals and a significant barrier for novices. 

Intelligent music production aims to develop systems for automating aspects of audio engineering~\cite{de2019intelligent}. 
Early approaches employed rule-based systems, using hand-engineered rules based on best practices~\cite{de2017ten}.
These system often generated outputs that satisfied certain assumptions or utilized well established conventions. 
However, the inability to construct sufficient sophisticated rule bases has motivated machine learning approaches, which instead learn from data without assuming a limited or fixed set of rules~\cite{scott2011automatic, moffat2019machine, martinez2021deep, steinmetz2021automatic}. 
Nevertheless, these systems still lacked the ability to adapt based on user input, which is critical to the context-dependent nature of music production~\cite{lefford2021context}.

\begin{figure}[t]
    \centering
    \vspace{-0.25cm}
    \includegraphics[width=0.97\linewidth, trim={1.1cm 0.15cm 0.4cm 0.2cm},clip]{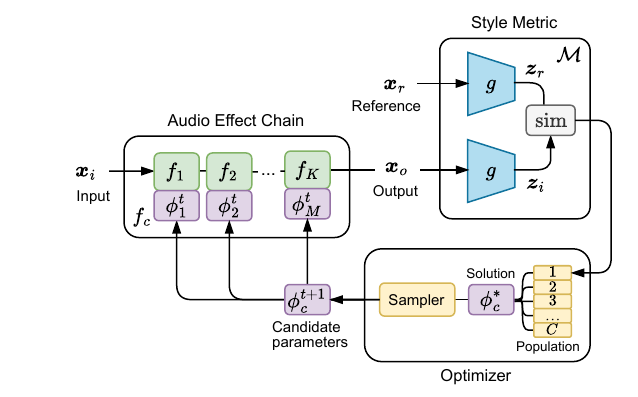}
    \vspace{-0.25cm}
    \caption{Style transfer with Inference-Time Optimization enables audio production style transfer through control of arbitrary audio effects. It employs a pretrained audio representation as a similarity metric, which is then optimized by searching the control parameter space of audio effects.}
    \label{fig:main-figure}
    \vspace{-0.3cm}
\end{figure}

To address the context-dependent nature of this task and enable greater user control, \emph{audio production style transfer} has been proposed~\cite{sheng2019feature, mimilakis2020one}.
These systems rely on a reference recording and attempt to map elements of the audio production style from the reference onto the input. 
These systems either directly process the audio signal~\cite{koo2021reverb, koo2022end, koo2023music} or estimate parameters for audio effects~\cite{sheng2019feature, mimilakis2020one, steinmetz2022style, peladeau2024blind}. 
While direct transformation methods are powerful, they may introduce artifacts and lack grounding in traditional audio tools. 
Similarly, recent text-to-audio generation models also enable editing capabilities~\cite{wang2023audit, han2023instructme}, but suffer from the same limitations. 
On the other hand, parameter based methods enable efficient and controllable style transfer.
However, current systems are limited to a fixed chain of effects, and require the use of differentiable signal processing~\cite{engel2020ddsp}, or inefficient alternative differentiation strategies such as gradient approximation~\cite{ramirez2021differentiable} and neural proxies~\cite{steinmetz2021automatic}.

We propose a method to construct an audio production style transfer system that leverages inference-time optimization to facilitate real world applications.
Instead of training a network to perform style transfer directly, we perform style transfer via an optimization process at inference, as shown in \fref{fig:main-figure}.
We iteratively search the parameter space of an effect chain with our proposed metric that measures the similarity in audio production style between the output recording and the reference.
This approach enables the ability to control arbitrary audio effect chains,  including non-differentiable effects, opening up the potential to control real-world audio effects.
The contributions of our work are as follows:
\begin{itemize}
\setlength\itemsep{-0.1em}
\item A simple and scalable pretraining strategy for constructing an audio production style similarity metric through audio effect estimation, named AFx-Rep.
\item A system for audio production style transfer, ST-ITO, that optimizes the control parameters of arbitrary audio effects according to a similarity metric. 
\item An extension of the DeepAFx-ST system~\cite{steinmetz2022style} with the addition of differentiable distortion and reverberation, which forms a strong baseline.
\item A multi-task benchmark for evaluation of audio production style similarity metrics and audio production style transfer systems. 
\end{itemize}
We provide audio examples, and open source our datasets, benchmark, and code to facilitate reproducibility\footnote{\url{https://github.com/csteinmetz1/st-ito}}.

\section{Method}

In this work, we propose \textbf{ST-ITO}, Style Transfer with Inference-Time Optimization, a novel method for audio production style transfer that searches the parameter space of a set of audio effects to perform style transfer.
As shown in \fref{fig:main-figure}, our system features three main components: an audio effect chain that processes an input recording, an audio production style similarity metric, composed of pretrained encoder and a similarity measure, and an optimizer that is used to find control parameters. 
This enables style transfer by finding a configuration of the audio effects that produce an output with attributes of the reference style. 

Our approach provides a number of benefits as compared to previous audio production style transfer systems. 
First, it enables the control of arbitrary audio effects, even those that have not been seen during training.
Unlike existing systems that train with a set of fixed effects ``in-the-loop'', our approach enables adaptation to new effects at inference.
Furthermore, our method removes requirements of previous systems.
This includes removing the need for differentiable audio effects or alternative differentiation strategies, which can often be slow and difficult to train~\cite{steinmetz2022style}.
Finally, we provide further flexibility and control by enabling the addition or removal of audio effects at inference without re-training of the base model.

\subsection{Audio production style metric}\label{sec:metric}

A central aspect of our approach is the development of an audio production style metric $\mathcal{M}(\vx_a, \vx_b)$. This metric measures the perceptual similarity in audio production between two recordings $\vx_a$ and $\vx_b$. 
As explained in \sref{sec:optim}, we optimize this metric by searching the parameter space of a set of audio effects to align the style of the processed recording with the reference. In general, production style relates to aspects of audio quality rather than the underlying content, including attributes such as dynamics, frequency balance, and the stereo field~\cite{vanka2024role}. 

\begin{figure}[t]
    \centering
    \includegraphics[width=\linewidth,trim={0.6cm 0.0cm 0.5cm 0.2cm},clip]{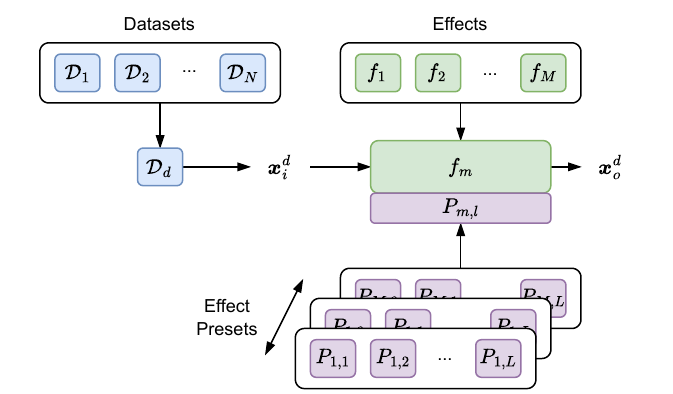}
    \caption{Self-supervised training for the pretext task where an audio signal $\vx_i^d \sim \mathcal{D}_d$ is sampled randomly from one of $N$ datasets and then processed by a randomly sampled audio effect $E_m$ with an associated randomly sampled parameter preset $P_{m,l}$ to produce an output signal $\vx_0^d$.}
    \vspace{-0.4cm}
    \label{fig:pretext-generation}
\end{figure}

\dotsubsec{Pretrained audio representations}
There is a growing body of work in general purpose representations of audio signals~\cite{turian2022hear}, popular approaches include CLAP~\cite{wu2023large} and BEATs~\cite{chen2023beats}. 
These representations capture relevant attributes to facilitate downstream tasks such as detection and classification of sound sources and events. 
While it may be possible to directly adapt one of these representations for our task, evidence suggests they are not always sensitive to audio effect transformations~\cite{hawley2023leveraging}.
We provide further evidence for this in \sref{sec:results}.
This motivates us to develop a method to produce our own audio representation that is more sensitive to audio effect transformations.


\dotsubsec{Self-supervised pretext task} 
We propose a simple and scalable self-supervised pretext task to construct an audio representation for our task without human annotated data.
To encourage the encoder to extract features related to audio effects we employ an audio effect classification task composed of two parts. 
The model predicts both which effect has been applied and the associated preset. 

As shown in \fref{fig:pretext-generation}, we generate training examples using $N$ audio datasets, a set of $M$ audio effects, and $L$ associated parameter presets for each effect. 
To generate a training example, a dataset $\mathcal{D}_d$ is selected at random from the set of datasets. 
Then one audio sample is selected from this dataset, which will form the input recording $\vx_i^d$. Next, we sample an audio effect $f_m$ and a random associated preset $P_{m,l}$ to configure the effect. 
Then we process the input with this effect to produce an output signal $\vx_o^d$. 

\begin{figure}[t!]
    \centering
    \includegraphics[width=0.9\linewidth,trim={1.5cm 0.1cm 0.0cm 0.2cm},clip]{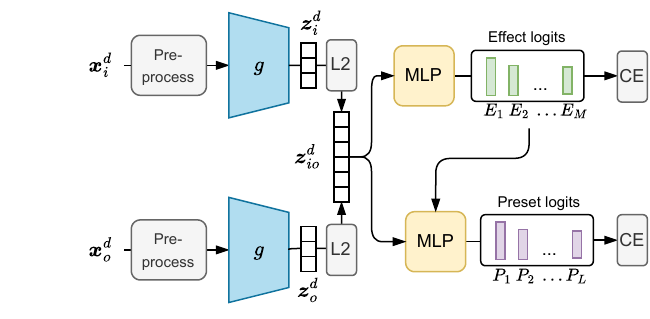}
    \caption{Representation learning via the pretext task where the input $\vx_i^d$ and output $\vx_o^d$ are processed by the encoder $g(\cdot)$ to produce embeddings. These embeddings are fed to a pair of MLP classifiers trained via cross-entropy that predict the effect class and preset class. 
    }
    \vspace{-0.4cm}
    \label{fig:pretext-prediction}
\end{figure}

Training is shown in \fref{fig:pretext-prediction} where the encoder $g(\cdot)$ extracts embeddings $\vz_i$ and $\vz_r$ from the input and output. 
These embeddings are concatenated and fed to a multi-layer perceptron (MLP) that estimates the effect applied. A second MLP takes the effect logits as well as the embeddings to estimate the preset. 
After pretraining we discard the prediction heads and use the encoder $g(\cdot)$, which we refer to as AFx-Rep, to produce embeddings.

This multi-task formulation encourages the encoder to extract features not only about effects but also subtleties between different configurations of the same effect, which is important for style transfer.
Our approach does not enforce invariance to the content, but can leverage any audio, not only unprocessed or effect normalized audio~\cite{martinez2022FxNormAutomix}, required by previous work~\cite{koo2023music}. This allows us to further scale the training dataset size. In addition, since this audio may already contain other processing, our model is exposed to a wide range of effects beyond those we apply.

\subsection{Inference-time optimization}\label{sec:optim}

To perform style transfer we begin with input $\vx_i$ and reference $\vx_r$ recordings. 
We assume the input has minimal processing, as our system does not remove effects~\cite{imort2022distortion, rice2023general}.
Then, we require the user to provide an appropriate chain of audio effects to be controlled. 
This chain can be represented as the composition of $K$ audio effects where each effect is represented by a function \(f_k\) parameterized by a control vector \(\phi_k\). The output \(\vx_o\) is obtained by sequentially applying these functions to the input, resulting in 
\begin{equation}\label{eq:general}
    \vx_o = f_K(f_{K-1}(\dots f_2(f_1(\vx_i; \phi_{1}); \phi_{2}) \dots ; \phi_{{K-1}}); \phi_{K}).
\end{equation}
For convenience, we represent this chain as a single function \(\vx_o = f_c(\vx_i; \phi_c)\), where \(\phi_c = [\phi_1, \phi_2, \ldots, \phi_K]\) concatenates all effect parameters into one vector.
While our method supports arbitrarily complex effect chains, we consider only series connections. 

In our setup, we perform style transfer through an optimization process via the maximization of a similarity between the output of our composite audio effect function and the reference signal given by
\begin{equation}
\max_{\phi_c} \ \ \textrm{sim} \Big( g(f_c(\vx_i; \phi_c)), g(\vx_r) \Big),
\end{equation}
where \(g(\cdot)\) denotes our audio representation, transforming audio signals into a feature space where audio production similarity is assessed. For the reference signal \(\vx_r\), the feature representation is \(\vz_r = g(\vx_r)\).
The optimization process initiates with a predefined set of control parameters, \(\phi_0\), and iteratively refines this estimate to enhance the similarity measure. At each step, candidate solutions are generated and evaluated based on their performance in mirroring the reference features, \(\vz_r\). This performance is quantified by the cosine similarity measure,
\begin{equation}
\textrm{sim}(\vz_i, \vz_r) = \frac{\vz_i \cdot \vz_r}{\max(\| \vz_i \| \| \vz_r \|, \epsilon)},
\end{equation}
where \(\vz_i = g(f_c(\vx_i; \phi_c))\) is the feature vector of the processed input signal, \(\cdot\) represents the dot product, \(\|\cdot\|\) denotes the Euclidean norm, and \(\epsilon\) is a small constant ensuring numerical stability, avoiding division by zero.


\section{Experimental details}

\subsection{Pretraining}
We employ the PANNs architecture~\cite{kong2020panns} as a convolutional backbone. 
While initial testing indicated that more modern architectures such as HTS-AT~\cite{chen2022hts} performed comparably, we found that PANNs was more efficient at inference.
To enable the encoder to capture stereo information we produce separate embeddings for the mid and side signals, concatenating them into a single embedding, applying L2 normalization to each embedding before concatenation. 

We train the encoder following the pretext task described in \sref{sec:metric}. 
We use seven publicly available audio datasets to cover a diverse range of audio content across music, speech, singing voice, and instruments. 
These datasets include MTG-Jamendo~\cite{bogdanov2019mtg}, ENST-Drums~\cite{gillet2006enst}, URSing~\cite{bochen2022ursing}, FSD50k~\cite{fonseca2021fsd50k},
Librispeech~\cite{panayotov2015librispeech},
Medley-solos-db~\cite{lostanlen2018medley}, and
GuitarSet~\cite{xi2018guitarset}.
To construct our set of audio effects we use 63 open source or freely available VST3 audio plugins compiled for Linux. 
These VSTs cover a wide range of effects including reverberation, dynamic range processing, equalization, distortion, modulation effects, and more.
We use \texttt{pedalboard}~\cite{sobot_peter_2023_7817838} to load plugins and apply effects to audio signals.

We generate unique presets for each effect by randomly sampling 1000 parameter configurations and processing a random audio recording with each configuration. 
Then we extract MFCCs and perform K-means clustering ($K=10$), with each cluster representing perceptually diverse parameter configurations. 
We then randomly select one configuration from each cluster to act as a preset. 

While training with on-the-fly data generation is possible, we found running VSTs during training caused a significant bottleneck. 
We opted for offline data generation, where we generated 20,000 examples of length 524288 samples ($\approx 11$\,sec at $f_s = 48$\,kHz) from each dataset with randomly applied effects and presets.
This corresponds to approximately 60 hours of audio content. 
We further increase diversity during training by taking different random crops of the pre-processed input and output segments, as well as applying random gain adjustments $[-32\,\textrm{dB}, 0\,\textrm{dB}]$. 

We perform pretraining for 1\,M steps with a batch size of 32 using the Adam optimizer. We use an initial learning rate of 1e-4, lowering the learning rate by a factor of 10 at 85\% and 95\% through training.  
We pre-process spectrogram inputs to the encoder by computing log-melspectrograms with window size of 2048 and hop size of 512. 
We clip the magnitudes between -80 and 40\,dB and scale the final spectrogram between -1 and 1. 

\subsection{Inference-time optimization}

To enable control of arbitrary effects, we employ a gradient-free optimizer as opposed to commonly used gradient-based solutions. 
While any gradient-free optimizer can be used in our system, we opt to use Covariance Matrix Adaptation Evolution Strategy (CMA-ES)~\cite{hansen1996adapting} since it has been shown to work well in spaces with similar dimensionality to the audio effect chain control parameter space ($\approx$100) and is a relatively scalable method. 
After initial hyperparameter tuning, we use a population size of 64 and a maximum of 25 optimization steps. The $\sigma$ hyperparameter is initialized to $0.3$ and we use a fixed initialization of all parameters of $0.5$ scaled in the range $[0, 1]$.  
We use early stopping, halting optimization after 10 steps of improvement less than 0.1. 

Unless otherwise specified, we use AFx-Rep as the encoder in our similarity metric, and we consider control of two different audio effect chains. 
The first features five VST audio effects including distortion (TubeScreamer), parametric equalizer (ZamEQ2), dynamic range compressor (ZamCompX2), feedback delay (ZamDelay), and artificial reverberation (TAL-Reverb-4), resulting in a total of 73 parameters. 
The second chain features unseen audio effects internal to \texttt{pedalboard}, including distortion, dynamic range compression, parametric equalizer, delay, and artificial reverberation, resulting in 36 parameters. 

\section{Benchmark}

\subsection{Audio production style metrics}
\dotsubsec{Zero-shot style classification}
We adapt the style classification task from \cite{steinmetz2022style}, which contains five different audio production styles using equalization and dynamic range compression: telephone (TL), bright (BR), warm (WM), broadcast (BC), neutral (NT). 
Training examples are generated by applying these style presets to speech from DAPS~\cite{mysore2014can} and music from MUSDB18~\cite{rafii2017musdb18}. 
To make the task more challenging and similar to the inference-time optimization use-case, we adapt the original task to the zero-shot case~\cite{wolters2020study}.
To do so, a query is constructed by sampling a random audio example to be classified as one of the five styles. 
Then other examples from each of the five styles are sampled randomly to form prototype classes. 
A representation of the query and each of the five prototypes is generated and a prediction is made by measuring the cosine similarity between the query and each of the prototypes. The class of the prototype with the highest similarity to the query forms the prediction. 

\dotsubsec{Style retrieval}
While the zero-shot style classification task evaluates the ability of a representation to differentiate among different styles, it considers only two basic effects and focuses on comparing significant differences. 
In order to more effectively evaluate the behavior of representations in a scenario similar to style transfer we designed a style retrieval task. 
In this task, a query style is produced by applying $N$ effects with random parameters to an audio recording. 
A retrieval set is generated by processing $M$ other recordings with differing random effect chains. 
One recording with differing content but the same effect chain as the query is included in the retrieval set. 
Similar to the zero-shot task, we measure the cosine similarity between the query and each of the items in the retrieval set. 
We can make the task more or less difficult by varying both the number of effects $N$ in each style and the size of the retrieval set $M$.
We source unseen audio examples for speech (DAPS~\cite{mysore2014can}), guitar (IDMT-SMT-Guitar~\cite{kehling2014automatic}), vocals (VocalSet~\cite{wilkins2018vocalset}), and drums (IDMT-SMT-Drums~\cite{dittmar2014real}).

\dotsubsec{Baselines}
We consider signal processing approaches, such as MFCCs and MIR features~\cite{man2014analysis}, as well as pretrained general purpose audio representations including VGGish~\cite{hershey2017cnn}, WAV2CLIP~\cite{wu2022wav2clip}, wav2vec2.0~\cite{baevski2020wav2vec}, CLAP~\cite{wu2023large}, as well as BEATs~\cite{chen2023beats}. 
We also compare against audio effect specific models including FX-Encoder~\cite{koo2023music} and the DeepAFx-ST encoder~\cite{steinmetz2022style}.

\setlength{\tabcolsep}{0.4em}
\renewcommand{\arraystretch}{0.85}
\begin{table}[t]
    \centering
    \begin{tabular}{l c c c c c c c c c c c c} \toprule
                & \multicolumn{5}{c}{Styles} & \\ \cmidrule(lr){2-6}
Representation & TL & BR & WM & BC & NT & AVG \\ \midrule 
MFCCs & \textbf{1.00} & 0.82 & 0.64 & 0.74 & 0.48 & 0.74 \\ 
MIR Feats. & 0.76 & 0.64 & 0.61 & 0.58 & 0.32 & 0.58 \\ \midrule
CLAP & 0.72 & 0.60 & 0.51 & 0.57 & 0.41 & 0.56 \\ 
Wav2Vec2 & 0.40 & 0.33 & 0.28 & 0.35 & 0.34 & 0.34 \\ 
Wav2Clip & 0.76 & 0.48 & 0.60 & 0.49 & 0.51 & 0.57 \\ 
VGGish & 0.47 & 0.58 & 0.43 & 0.61 & 0.43 & 0.50 \\ 
BEATs & 0.94 & 0.51 & 0.57 & 0.50 & 0.45 & 0.59 \\ \midrule
FX Encoder & 0.96 & 0.94 & 0.29 & 0.70 & 0.54 & 0.69 \\ 
DeepAFx-ST & \textbf{1.00} & 0.93 & 0.67 & 0.78 & 0.42 & 0.76 \\ 
DeepAFx-ST+ & \textbf{1.00} & 0.97 & 0.71 & 0.79 & 0.41 & 0.78 \\  \midrule
AFx-Rep (ours) & \textbf{1.00} & \textbf{1.00} & \textbf{0.88} & \textbf{0.85} & \textbf{0.59} & \textbf{0.86} \\ 
    \bottomrule
    \end{tabular}
    \caption{Zero-shot style classification accuracy over 200 trials for music and speech across five unique styles.}
    \vspace{-0.4cm}
    \label{tab:zero-shot}
\end{table}

\subsection{Audio production style transfer}
\dotsubsec{Parameter estimation}
To demonstrate the ability of our proposed approach to control a wide range of effects we design a parameter estimation task. 
We initialize an audio effect and set a target value for one parameter, processing a random audio signal to generate a reference. 
Then we sample another random recording to use as the input. 
We then run the optimization using each audio representation in our metric. 
To achieve accurate style transfer a system should estimate a control parameter with a similar, but not necessarily identical value as the reference. 
We report both the mean squared error (MSE) and the correlation coefficient $\rho$ of estimated parameters.
We consider six VST effects as well as six unseen effects from \texttt{pedalboard}.

\dotsubsec{Real-world style transfer}
While the parameter estimation task can demonstrate the ability of a style transfer system to control a singular audio effect, it does not capture the ability of the system to perform in a real-world scenario. 
In many cases multiple effects will be present at the same time, making the task more challenging. 
To evaluate this scenario we created six different audio production styles by constructing realistic audio effect chains of varying complexity in a digital audio workstation. 
These styles range from simple lowpass and highpass filtering to a complete channel strip featuring equalization, distortion, compression, delay, and reverberation.
We then applied these styles to a range of audio content including speech, singing voice, and full music tracks.  
Each style transfer system is then tasked with transforming the unprocessed audio from one of these content types to the stylized version of a different recording containing the same kind of content. 

\begin{figure}[t]
    \centering
    \vspace{-0.3cm}
    \includegraphics[width=\linewidth,trim={0.0cm 0.4cm 0.7cm 0.5cm},clip]{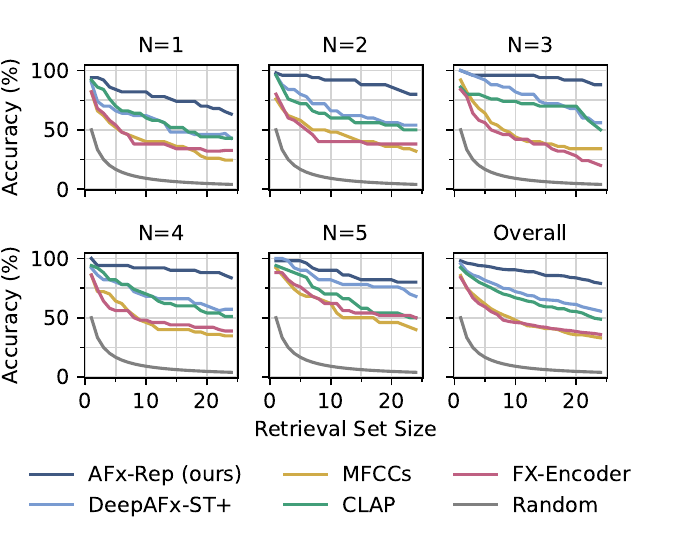}
    \vspace{-0.5cm}
    \caption{Accuracy for the style retrieval task using different audio representations across multiple source types with varying number of audio effects ($N$) and retrieval set size.}
    \label{fig:style-retrieval}
    \vspace{-0.4cm}
\end{figure}

\dotsubsec{Baselines}
We compare our proposed style transfer system to both deep learning and signal processing solutions. 
We use a rule-based approach from previous work that includes an FIR matching equalization filter and a simple hill climbing-based dynamic range compressor~\cite{steinmetz2022style}. 
We construct a strong deep learning baseline by extending DeepAFx-ST with differentiable reverberation~\cite{steinmetz2021filtered} and distortion~\cite{colonel2022reverse} effects, using \texttt{dasp-pytorch}\footnote{\url{https://github.com/csteinmetz1/dasp-pytorch}}.
We call this approach DeepAFx-ST+ and we train this model following the approach from the original work, but using the same datasets used to train our audio representation.


\setlength{\tabcolsep}{0.3em}
\renewcommand{\arraystretch}{0.9}
\begin{table}[t]
\vspace{-0.2cm}
    \centering
    \begin{tabular}{l c c c c  } \toprule
           & \multicolumn{2}{c}{MSE ($\downarrow$)} & \multicolumn{2}{c}{$\rho$ ($\uparrow$)} \\ \cmidrule(lr){2-3} \cmidrule(lr){4-5}
           Effect (Parameter)     &  CLAP & AFx-Rep & CLAP & AFx-Rep \\ \midrule
RoughRider (sensit) & 0.183 &\textbf{0.084} & 0.300 &\textbf{0.705} \\ 
DPlate (decay) & 0.141 &\textbf{0.025} & 0.610 &\textbf{0.945} \\ 
3BandEQ (high\_) & 0.033 &\textbf{0.026} & 0.876 &\textbf{0.919} \\ 
MaGigaverb (size) & 0.018 &\textbf{0.012} & 0.949 &\textbf{0.969} \\ 
MetalTone (dist) & 0.155 &\textbf{0.040} & 0.509 &\textbf{0.862} \\ 
TAL-Chorus (wet) & 0.097 &\textbf{0.014} & 0.654 &\textbf{0.953} \\ \midrule
*Chorus (mix) & \textbf{0.164} & 0.172 & 0.300 &\textbf{0.408} \\ 
*Reverb (room\_) & 0.048 &\textbf{0.013} & 0.822 &\textbf{0.955} \\ 
*Delay (mix) & 0.117 &\textbf{0.052} & 0.591 &\textbf{0.815} \\ 
*Distortion (drive) & 0.023 &\textbf{0.005} & 0.852 &\textbf{0.944} \\ 
*Compressor (thresh) & 0.134 &\textbf{0.096} & 0.518 &\textbf{0.678} \\
*ParametricEQ (low\_s) & 0.110 &\textbf{0.031} & 0.727 &\textbf{0.931} \\ 
\bottomrule
    \end{tabular}
    \caption{Parameter estimation with ST-ITO using CLAP and our proposed AFx-Rep. We report the mean squared error (MSE) and correlation coefficient ($\rho)$ of the estimated parameters across 4 different settings and 3 trials per effect. Audio effects not seen during pretraining are denoted by $*$.}
    \label{tab:parameter-estimation}
    \vspace{-0.2cm}
\end{table}

\section{Results}\label{sec:results}

\dotsubsec{Zero-shot style classification}
We evaluate the pretrained representations across ten trials for each of the five different styles. 
The class-wise and overall accuracy is reported in \tref{tab:zero-shot}.
First, we find MFCC based features perform better than expected, with high accuracy on the telephone (TL), bright (BR), and warm (WM) styles. However, performance is worse on broadcast (BC) and neutral (NT), likely because identifying these styles requires paying attention to dynamics.
The MIR features do not achieve comparable performance. 
All of the general purpose audio representations perform worse than MFCCs on this task, with CLAP and BEATs appearing to perform best among them, but with an average accuracy 15 points lower. 
This confirms the hypothesis that general purpose representations fail to capture information about audio effects. 
FX-Encoder and DeepAFx-ST(+) perform better than the other pretrained models, with DeepAFX-ST variants outperforming MFCCs. 
Overall, we find that our proposed model, AFx-Rep, performs best in this task.

\dotsubsec{Style retrieval}
We report the accuracy for a subset of methods in style retrieval as shown in \fref{fig:style-retrieval}. 
We plot performance across differing number of effects $N$ that constitute a style, as well as the size of the retrieval set, shown on the x-axis. 
As expected, for all scenarios, as the retrieval set grows the classification performance drops.
While all methods are better than random guessing, we observe that MFCCs and FX-Encoder appear to perform worse. 
They are followed by CLAP and then the encoder from DeepAFx-ST+, which slightly outperforms CLAP. 
Finally, our proposed AFx-Rep model performs best across all scenarios, indicating its superior ability to capture elements related to audio production style. 


\dotsubsec{Parameter estimation}
In \tref{tab:parameter-estimation} we report the mean squared error (MSE) and correlation coefficient ($\rho$) in parameter estimation using ST-ITO with either CLAP or our proposed AFx-Rep model.
In nearly all cases our AFx-Rep model functions as a superior similarity metric, achieving lower MSE and a higher correlation coefficient, with the exception of the MSE in Chorus, which appears to be challenging for both models. 
This demonstrates the ability of our approach to control a wide range of real-world audio effects, including effects not seen during retraining. These results reinforce the importance of an audio representation sensitive to audio effects, such as our proposed AFx-Rep.

\begin{figure}[t]
    \centering
    \includegraphics[width=1\linewidth,trim={0.0cm 0.4cm 0.5cm 0.2cm},clip]{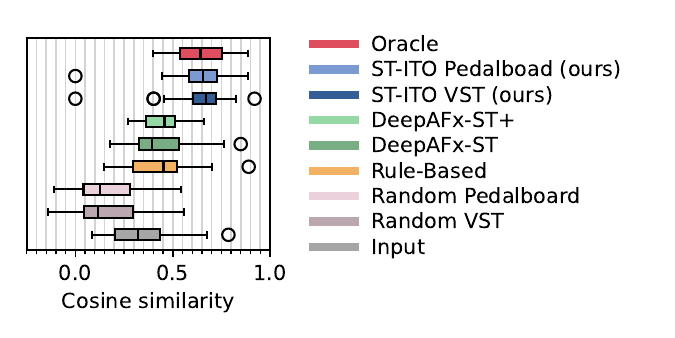}
    \vspace{-0.2cm}
    \caption{AFx-Rep similarity in real-world style transfer.}
    \vspace{-0.4cm}
    \label{fig:objective}
\end{figure}

\dotsubsec{Real world style transfer}
We report the similarity from our metric using AFx-Rep across 56 style transfer trials (\fref{fig:objective}).
The Input processed with an audio effect chain identical to the reference is also evaluated, which we refer to as Oracle.
Note that the Oracle may not achieve effective style transfer as the starting point of the Input may require a different parameter configuration to match the reference.
The random configuration of VSTs and \texttt{pedalboard} effect perform worse, and are followed by the Input, which features no processing. 
DeepAFx-ST, DeepAFx-ST+, and the Rule-Based system appear to perform similarly to each other, but better than Input.
Variants of ST-ITO, one using VSTs and the other using unseen \texttt{pedalboard} effects, both outperform the rest, and are on par with the Oracle.
This indicates the ability of our approach to optimize our metric,  
however, it is difficult to make conclusions about style transfer performance using this evaluation alone.

\begin{figure}[t]
    \centering
    \includegraphics[width=0.90\linewidth,trim={1.2cm 1.1cm 1.2cm 1.3cm},clip]{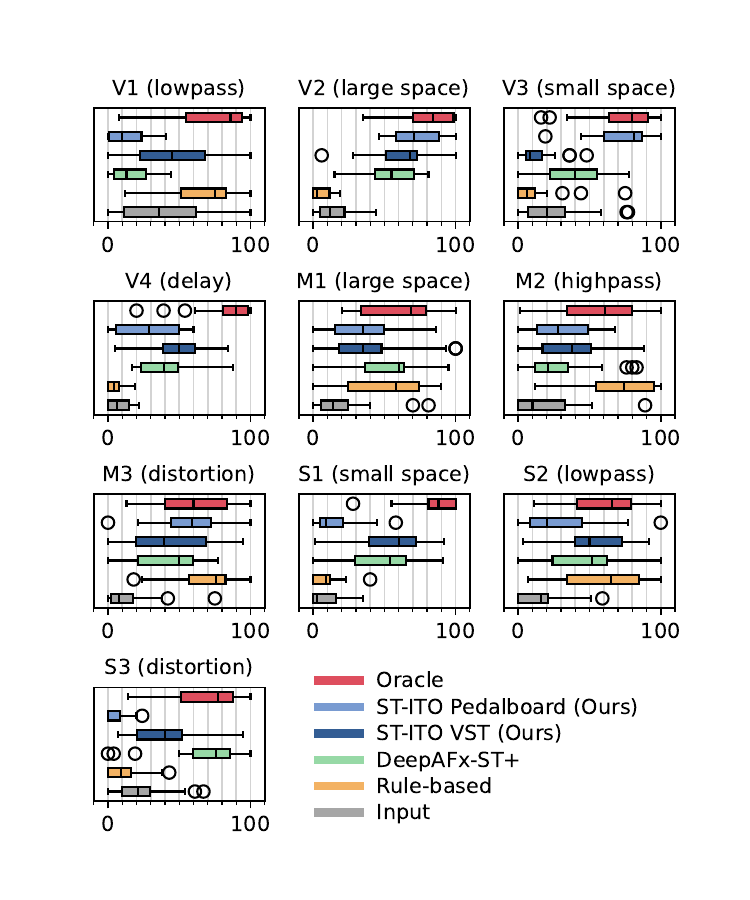}
    \vspace{-0.3cm}
    \caption{Subjective scores from $N=23$ participants across vocals (V), music (M), and speech (S) examples.}
    \vspace{-0.35cm}
    \label{fig:listening-test}
\end{figure}

\dotsubsec{Subjective listening study}
We recruited 23 participants with experience in audio engineering. They were tasked with evaluating style transfer systems on real-world test scenarios in a multiple stimulus listening study. 
Listeners were asked to provide a score from 0 to 100 for each stimulus to indicate its similarity to the reference, considering only the style and not the underlying content.
In addition, we also included the unprocessed Input and the Oracle. 
Due to the subjectivity of this task, evaluators may not rate Input the lowest and Oracle the highest.
We selected ten test cases across the real-world scenarios including vocals (V), music (M), and speech (S) as shown in \fref{fig:listening-test}.

Overall, listeners found the Oracle most similar to the reference and the Input the least similar, as expected. 
However, there is variation in the score assigned to both, indicating some disagreement. 
For simple styles, such as V1 (lowpass) and M2 (highpass), we found the Rule-Based system worked well, even surpassing style transfer systems. 
However, in cases with multiple effects, the Rule-Based system does not work well, as in V2 (large space), V3 (small space), V4 (delay), S1 (small space), and S3 (distortion). 
Differences between ST-ITO and DeepAFx-ST+ are harder to discern. 
Our method outperforms in some cases, such as V2 (large space), V3 (small space), and V4 (delay), yet in in other cases there is no clear difference. 
We conclude that our approach is capable of style transfer at least on par with the enhanced DeepAFx-ST+, and does so controlling a chain of unseen VST audio effects, which is not possible with other approaches.

\section{Discussion}
While ST-ITO enables control of arbitrary audio effects and adapts to new effects at inference, it has some limitations. 
In the current formulation, our system requires an appropriate audio effect chain be provided. Future work could consider automatically constructing this audio processing graph as in blind estimation~\cite{lee2023blind}.
Furthermore, while our method does not require training ``in-the-loop'' with audio effects during representation learning, we must process many variants of the recording through the audio effect chain during style transfer.
This leads to significantly longer inference times ($\approx 1$\,min) as compared to networks that estimate parameters directly ($\approx 1$\,sec). 
Future work could consider the design of more efficient optimizers through meta-learning by training an optimizer for a particular effect chain~\cite{casebeer2022meta}.
Finally, we have found that the current system does not work well for challenging style transfer applications, such as guitar tone matching.

\section{Conclusion}

In this work, we introduced ST-ITO, Style Transfer with Inference-Time Optimization. 
Unlike previous style transfer systems, \mbox{ST-ITO} searches the parameter space of any audio effect chain at inference, enabling control of arbitrary effect chains, including those with non-differentiable effects. Our methodology leverages a self-supervised audio production style metric and a gradient-free optimizer. We developed a set of benchmarks to evaluate both audio production style representations and style transfer systems. Results from this set of benchmarks indicate that our approach not only better captures details related to audio production style, but also provides enhanced flexibility and expressiveness in audio production style transfer.

\clearpage

\section{Acknowledgments}
Supported by EPSRC UKRI CDT in AI+Music (Grant no. EP/S022694/1). EB is supported by RAEng/Leverhulme Trust research fellowship LTRF2223-19-106.

\bibliography{ISMIRtemplate}

\end{document}